\documentclass[twocolumn]{revtex4-1}
\usepackage{graphicx}
\usepackage{color}
\newcommand{\ket}[1]{\left| #1 \right\rangle}
\newcommand{\bra}[1]{\left\langle #1 \right|}

\newcommand{\ain}{a_{{\rm in}}}
\newcommand{\Noise}{N_{{\rm in}}}
\newcommand{\tNoise}{\tilde{N}_{{\rm in}}}

\newcommand{\cin}{c_{{\rm in}}}

\newcommand{\aout}{a_{{\rm out}}}

\newcommand{\adout}{a^{\dagger}_{{\rm out}}}

\newcommand{\bdin}{b^{\dagger}_{{\rm in}}}
\newcommand{\proj}[1]{| #1\rangle\!\langle #1 |}

\renewcommand{\Re}{\mathrm{Re}}
\newcommand{\expect}[1]{\left\langle#1\right\rangle}
\newcommand{\eea}{\end{eqnarray}}
\newcommand{\bea}{\begin{eqnarray}}
\newcommand{\ee}{\end{equation}}
\newcommand{\be}{\begin{equation}}

\newcommand{\ido}{\int_{0}^\infty\!d\omega\,}
\newcommand{\tF}{\tilde{F}}
\newcommand{\ddt}{\frac{d}{dt}}

\begin{document}
\title{The Heisenberg picture of photodetection}

\author{Saumya Biswas and S.J. van Enk}

\affiliation{Department of Physics and
Oregon Center for Optical, Molecular \& Quantum Sciences\\
University of Oregon, Eugene, OR 97403}

\begin{abstract}
We construct a class of Hamiltonians that describe the photodetection process from beginning to end.
Our Hamiltonians describe the
creation of a photon, how the photon travels to an absorber (such as a molecule), how the molecule absorbs the photon, and how the molecule after  irreversibly changing its configuration triggers an amplification process---at a wavelength that may be very different from the photon's wavelength---thus producing a macroscopic signal. We use a simple prototype Hamiltonian to describe the single-photon detection process analytically in the Heisenberg picture, which neatly separates desirable from undesirable effects. 
Extensions to more complicated Hamiltonians are pointed out.
\end{abstract}
\maketitle
\section{Introduction}

We may distinguish two traditional types of photodetection theory. The first tries to determine what quantum field observable is measured when photoelectrons are produced by photoabsorption 
and the photoelectrons are subsequently detected and/or counted. This sort of theory is exemplified by the classic papers
Refs.~\cite{glauber1963,kelley1964}, which take as starting point the Hamiltonian describing the interaction between photons and an electric dipole, but which do not describe the remainder of the detection process quantum mechanically.
The second type of theory models actual photodetectors
phenomenologically, taking  great care to model the many mechanisms involved in converting the initial photon energy to the final macroscopic current. This type of theory is exemplified by recent work on the superconducting nanowire detector \cite{yang2007,renema2014,frasca2019}.

Neither of the above types of photodetection theory establishes {\em fundamental} limits of photodetection, that is, platform-independent limitations arising from the laws of physics on, e.g., single-photon detection efficiency, dark count rates, time and wavelength resolution and tradeoffs between these figures-of-merit. 

For just this purpose---finding fundamental limits on photodetectors---a third type of theory has been developed in recent years. Here the aim is to develop fully quantum-mechanical and sufficiently realistic models that include all stages of the photodetection process, including the crucial amplification process \cite{vanenk2017,young2018,young2018b,yang2019,yang2019b,propp2019,proppb,propp2020}.
The point of this paper is to continue this recent work and present a quantum description of the processes involved in the detection of a single photon, especially the connection between the photoabsorption and  amplification processes.
Moreover, we perform our calculations  {\em in the Heisenberg picture}. That picture may not be the most intuitive---it may be easier to follow the trajectory of an excitation through the system in the Schr\"odinger picture---but it does have its merits. We mention two reasons here to
use this picture.

First, lower limits on noise accompanying quantum amplification are most easily derived in the Heisenberg picture. Caves \cite{caves1982}
studied linear amplification of electricmagnetic (EM) field amplitudes and formulated the problem in terms of the Heisenberg evolution of the annihilation operator of a single discrete EM field mode of the schematic form
\be
\aout=\sqrt{G}\ain + \Noise,
\ee
with $G>1$ the gain factor and $\Noise$ a noise term. The left-hand side here represents the annihilation operator for the mode to be amplified at the end of the amplification process, the operators on the right-hand side represent input operators, i.e., initial values just before the amplification process starts. Ideally, the  number of excitations in the output equals the  number of input excitations multiplied by $G$, and this would be the case if it were not for the noise term $\Noise$.

The commutator $[a,a^{\dagger}]=\openone$ has to be preserved, i.e., at any time $t$ we must have $[a(t),a^{\dagger}(t)]=\openone$ for the Heisenberg-picture operators $a(t)$ and $a^{\dagger}(t)$. This puts a restriction on the noise operator $\Noise$. In particular, it cannot be zero. For example, phase-insensitive amplification is obtained by setting 
\bea\Noise=\sqrt{G-1}\bdin,
\eea 
in terms of the creation operator of an additional discrete bosonic mode $b$. 
As is easily verified, the addition of that noise term preserves the commutator. Thermal excitations in the additional mode $b$ are amplified, too, by a factor of $G-1$; and even if mode $b$ starts in the vacuum state (at zero temperature) the fact that it is the mode's {\em creation} operator appearing here still causes noise. We refer to the lower limit on noise reached here as the Caves limit on linear amplification.

Second, in a recent paper \cite{propp2019} we showed that the first part of the photo detection process (the part preceding amplification, including absorption of the incident photon) can be described 
 compactly in the Heisenberg picture as well. 
The input-output relations for the annihilation operators (which now are continuous-mode operators indexed by frequencies $\omega$)  consist of two clearly distinct terms, one desired, the other undesired but inevitable so as to satisfy the commutator
 $[\aout(\omega),\adout(\omega')]=\delta(\omega-\omega')$. We can write
 \be
 \aout(\omega)=T(\omega)\ain(\omega) + \tNoise(\omega),
 \ee 
 where $T(\omega)$ is a complex transmission amplitude, with the physical meaning that a photon with frequency $\omega$ will survive the pre-amplification stage with probability $|T(\omega)|^2$ (we will encounter this interpretation in Eqs.~(\ref{Tom}--\ref{phiom})).
 Here the noise term is of the form
 \bea
 \tNoise(\omega)=R(\omega)\cin(\omega),
 \eea
 with $\cin(\omega)$ the annihilation operators for external bosonic modes at frequency $\omega$, and 
 \bea
 |R(\omega)|^2+|T(\omega)|^2=1
 \eea
 so as to preserve the commutator $[\aout(\omega),\adout(\omega')]$.
Once again, thermal excitations at arbitrary frequencies $\omega$ in the mode $\cin(\omega)$  contribute noise as soon as $R(\omega)\neq 0$.

There are several reasons for wishing to describe the whole photo-detection process with one Hamiltonian.
First, although in a recent paper \cite{proppb} it was shown that one can write down commutator-preserving (nonlinear) amplification relations that beat the above-mentioned Caves limit, no explicit Hamiltonians were considered there that may reach that improved limit. Here, we reach the same improved limit, but in a new way and with a (fairly simple) Hamiltonian.
Second, that same paper also noticed how one can 
formally express the idea that one can amplify at a frequency that differs substantially from the incoming photon frequency. We show here explicitly how that idea can be implemented, quite straightforwardly, by a Hamiltonian.

\section{A class of model Hamiltonians}
\subsection{Description}
We wish to represent the whole photo-detection process (including absorption of the photon and amplification) plus the generation of the photon to be detected, by a Hamiltonian.
We start with what seems to be a minimal model (various possible extensions of the model are discussed in the concluding Section).
There are 6 quantum systems in total; we have 3 {\em discrete} quantum systems $a$, $F$, and $c$ (with small Hilbert-space dimensions, which generate the photon, absorb the photon, and amplify the signal, respectively) and 3 {\em continuous-mode} quantum systems that connect the discrete systems and that are used to model irreversible processes (see Fig.~1). 

The continuous modes are modeled by bosonic mode operators $b(\omega_b), d(\omega_d), g(\omega_g)$ with $\omega_i=ck_i$ proportional to the  wave number $k_i$ (using just 1 spatial dimension, the $x$-axis) of the bosonic excitations of type $i=b,d,g$. 
When there is no confusion possible, e.g., when we integrate over all frequencies $\omega_i$, we will use the  symbol $\omega$ without subscript to denote those. Positive frequencies $\omega>0$ describe waves traveling from left to right (towards positive $x$), $\omega<0$ waves traveling from right to left.
 \begin{center}
\begin{figure}
\includegraphics[width=4in]{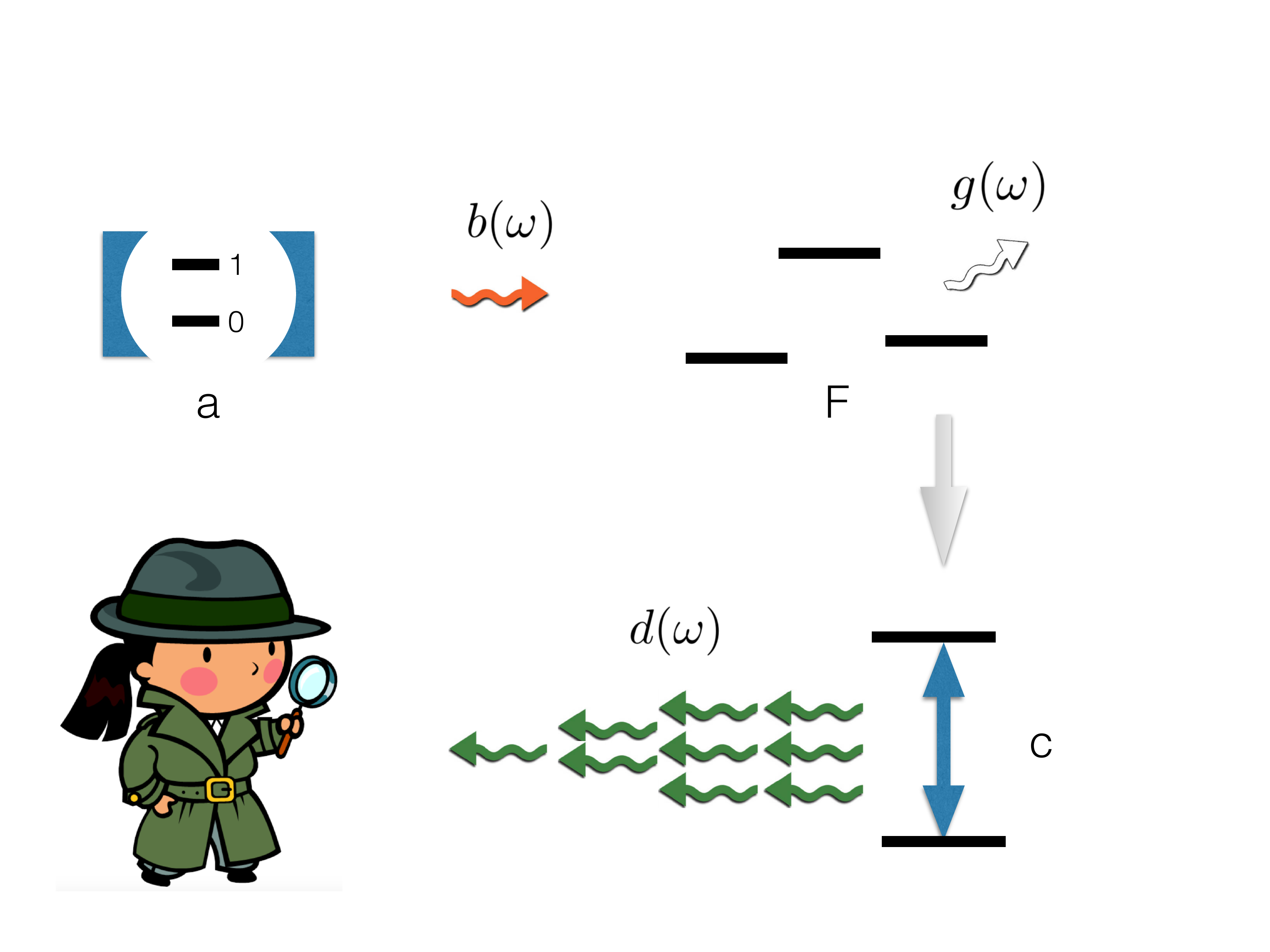}
\caption{From a single photon to a macroscopic signal: a cavity $a$ which contains one excitation generates one single-photon wavepacket (in a continuum $b(\omega)$). That photon is (resonantly) absorbed by a molecule $F$. The molecule may decay back to its initial state $\ket{F_0}$ or it may decay to a different state $\ket{F_2}$ by emitting a different photon (in a continuum $g(\omega)$) that escapes. In the state $\ket{F_2}$ the molecule's shape and/or dipole moment have changed. That physical change triggers an amplification process in another system $c$, which eventually reaches a steady state in which spontaneous decay is balanced by a ``classical'' drive, thus producing a stream  of fluorescence photons (in a continuum $d(\omega)$) that a classical (human) observer can observe. 
	The final macroscopic signal may be at a (very) different wavelength than that of the single photon.
}
\end{figure}
\end{center}
System $a$ is a cavity that contains a single excitation that leaks out into the continuum described by $b(\omega$), and this is
the single photon wave packet that we intend to detect. That single excitation couples to to first leg of a three-level $\Lambda$ system---a ``molecule'' which we denote by $F$ because it is the driving force behind amplification.  The photon may excite the molecule from its initial state, the lower level $\ket{F_0}$, to an excited state $\ket{F_1}$. That level could decay back to $\ket{F_0}$ or it could decay to the state $\ket{F_2}$ through coupling to the continuum $g(\omega)$ [the excitation in $g$ thus produced is assumed to escape; this is an irreversible transition]. When the molecule $F$ is in the state $\ket{F_2}$  the accompanying physical change in the molecule triggers an amplification process in system $c$. 

This aspect of the model mimics the mechanism used in the human eye: a retinal molecule changes its configuration from cis to trans,
and that change of shape in turn induces a conformational change in the protein the retinal binds to. Further changes in shapes of proteins then finally lead to a change in the charge distribution, that then can generate an electric signal (see \cite{hall2010}). 
This idea  can be exploited in a bioinspired photodetector \cite{zhou2009,leonard2019} where a chromophore molecule changes its shape upon absorbing a single photon and thereby changes its dipole moment, which then affects a (macroscopic) current.

Finally, the many excitations generated by $c$ leak out to the continuum mode $d(\omega)$. The macroscopic signal present in  continuum $d(\omega)$ is what we then (classically) observe (see Fig.~1).
\subsection{Hamiltonian}

We follow here the ideas of Gardiner \cite{gardiner1993,gardiner1985} (see also \cite{molmer2019}) for describing how the output of one quantum system may serve as the input for the next quantum system, without the latter acting back on the first system. This is done simply by setting the coupling to the right-to-left traveling waves equal to zero. That is, we only need the positive frequencies here.
The (electric) field operators describing fields that travel from left to right corresponding to the modes $b, d$, and $g$ are denoted by corresponding capital letters $B(x),D(x)$ and $G(x)$, and are expanded as
\bea
B(x,t)&=&\frac{1}{\sqrt{2\pi}}\ido b(\omega,t)\exp(i\omega x/c),\nonumber\\
D(x,t)&=&\frac{1}{\sqrt{2\pi}}\ido d(\omega,t)\exp(i\omega x/c),\nonumber\\
G(x,t)&=&\frac{1}{\sqrt{2\pi}}\ido g(\omega,t)\exp(i\omega x/c).
\eea
It is through these field operators together with their hermitian conjugates $B^{\dagger}(x), D^{\dagger}(x)$ and $G^{\dagger}(x)$ that the discrete systems interact (at their resp\label{key}ective locations on the $x$ axis) with the continuous modes, where we will make both Markov and rotating-wave approximations (RWA), as detailed below.

The Hamiltonian is of the following form
\bea\label{H}
H=H_a+H_{a-b}+ H_b\nonumber\\
+H_{b-F}+
H_F+H_{F-g}+H_g\nonumber\\
+H_{F-c}+H_c\nonumber\\
+H_{c-d}+H_d.
\eea
System $a$ is a single cavity mode with resonance frequency $\omega_a$ (which, in all generality, would be time-dependent), whose Hamiltonian we write as (setting $\hbar=1$ everywhere)
\be
H_a=\omega_aa^{\dagger}a.
\ee
The cavity mode is located at $x=-c\tau$
with $\tau$ the time delay between a signal (a photon) leaving system $a$ and interacting with system $F$ (which we take to be located at $x=0$).
The cavity mode is
coupled to the field $B(x=-c\tau,t)$  like so:
\be
H_{a-b}=i \sqrt{\kappa} [aB^{\dagger}(-c\tau,t)-B(-c\tau,t)a^{\dagger}],
\ee
with the field $B$ described by the Hamiltonian
(recall we leave out negative frequencies since they do not couple to the systems of interest)
\be
H_b=\ido \omega b^{\dagger}(\omega)b(\omega).
\ee
We can anticipate that the main terms contributing to the interaction are those at $\omega\approx\omega_a$. That is,
in the Heisenberg picture $a(t)\sim\exp(-i\omega_a t)$ and $b^{\dagger}(\omega,t)\sim\exp(+i\omega t)$ so that the main terms not averaging to zero over time come from $\omega\approx\omega_a$. In numerical simulations, we always transform to a rotating frame, i.e., we solve equations for the slowly-varying operators
$\exp(i\omega_a t)a(t)$ rather than $a(t)$, etcetera.

The next line of (\ref{H}) contains four Hamiltonians that describe how the $\Lambda$  system $F$ interacts with two different continua (namely, $b$ and $g$) at position $x=0$. 

The four terms are written as
\bea
H_{b-F}&=&i\sqrt{\gamma_1}[\ket{F_0}\bra{F_1}B^{\dagger}(x=0,t)-B(x=0,t)\ket{F_1}\bra{F_0}]\nonumber\\
H_F&=&\sum_{k=0,1,2} \omega_k \proj{F_k}
\nonumber\\
H_{F-g}&=&i\sqrt{\gamma_2}[\ket{F_1}\bra{F_2}G^{\dagger}(x=0,t)-G(x=0,t)\ket{F_2}\bra{F_1}]\nonumber\\
H_g&=&\ido \omega g^{\dagger}(\omega)g(\omega)
\eea
and we again can anticipate that the most important terms are those with 
$\omega\approx \omega_1-\omega_0:=\omega_{10}$ for the interaction between $F$ and $b(\omega)$ and for
$\omega\approx (\omega_1-\omega_2)$ for the interaction between $g(\omega)$ and $F$. An important parameter is the detuning of the photon from resonance with the molecular transition from $\ket{F_0}$ to $\ket{F_1}$,
\be
\delta=\omega_a-\omega_{10}.
\ee

The third line of (\ref{H})  is going to be crucial as it models the amplification process and how the $F$ system triggers it.
We construct the two Hamiltonians $H_{F-c}$ and $H_c$ in several steps. First,  assume we have a system $c$ that is driven by an external ``force'' $F$. That is, we assume the Hamiltonian for system $c$ contains a driving term proportional to a parameter $F$. For example, we may use a Hamiltonian that describes {\em electron shelving} \cite{nagourney1986,bergquist1986}, which is used as a method to perform quantum state measurements on ions: in one state the ion, driven by a laser, produces large amounts of fluorescence whereas in another state (from which there is no transition resonant with the laser) it remains dark.
The simple Hamiltonian is [in Section \ref{discuss} below we suggest several more involved examples of suitable Hamiltonians]:
\be
\tilde{H}_c(F)=2iF \cos(\omega_F t) (c-c^{\dagger}),
\ee
which in the RWA becomes
\be\label{HcF}
\tilde{H}_c(F)\approx iF  (\exp(i\omega_F t)c-\exp(-i\omega_F t)c^{\dagger}).
\ee
Alternatively, we could introduce yet another pair of bosonic mode operators $\alpha$ and $\alpha^{\dagger}$ to replace $F\exp(-i\omega_F t)$ and $F\exp(i\omega_F t)$, respectively, and add another Hamiltonian $H_\alpha=
\omega_F \alpha^{\dagger}\alpha$; the initial state of that mode would then be a coherent state with  amplitude $F$, i.e., an eigenstate of $\alpha$ with eigenvalue $F$ (which indeed can model a strong laser field). This would have the formal advantage of making our Hamiltonian completely independent of time.

Now in order to couple system $c$ to our quantum system $F$ we replace the parameter $F$ by the quantum operator
\be
\tF=\sum_{k=0,1,2}F_k\proj{F_k},
\ee
so that we replace
\be
\tilde{H}_c(F)\mapsto \tilde{H}_c(\tF).
\ee
Note we may apply this substitution trick to {\em any} Hamiltonian $\tilde{H}_c(F)$ that models amplification. 
For the simple example (\ref{HcF}) we have
\be
\tilde{H}_{F-c}
=i\tF(\exp(i\omega_F t)c
-\exp(i\omega_F t)c^{\dagger}).
\ee
For $\tilde{H}_c$ we assume the simple form
\be
\tilde{H}_c=\omega_c c^{\dagger}c,
\ee
appropriate for a bosonic mode $c$ (but we could also use a two-level atom).
An interaction proportional to a projector $\proj{F_2}$, implements the idea (mentioned above) that it is a physical property of the state $\ket{F_2}$ that triggers amplification.
For the RWA to apply all we need is that the driving frequency $\omega_F$ be close to the frequency $\omega_c$, {\em independent} of the frequencies $\omega_k$
for the three states of system $F$ and {\em independent}  of $\omega_a$, the frequency of the photon to be detected.
And so amplification happens at the frequency $\omega_F$, not at $\omega_a$. It may be important to be able to amplify at a different frequency so as to suppress thermal noise (which could lead to dark counts) by amplifying at a frequency $\omega_F$ such that (reinserting $\hbar$) $\hbar\omega_F\gg kT$.
Pre-amplification thermal excitations at a frequency $\omega_a$ [either thermal photons in the input mode $a$ or thermal fluctuations exciting the transition to level $\ket{F_1}$] are amplified, and so, too, should be suppressed by operating at a temperature $T$ such that  $\hbar\omega_a\gg kT$. (For example, this is how our eyes can detect optical photons at room temperature.)

We may move to a frame rotating at frequency $\omega_F$ and replace our first guesses for Hamiltonians $\tilde{H}_{F-c}$ and $\tilde{H}_c$ by
the final results 
\bea
H_{F-c}
&=&i\tF(c-c^{\dagger}),\nonumber\\
H_c&=&\Delta c^{\dagger}c,
\eea
with $\Delta=\omega_c-\omega_F$ the detuning from resonance.

Finally, for the fourth line in (\ref{H}) we stay in the same rotating frame and write
\bea
H_{c-d}&=&i \sqrt{\Gamma} [cD^{\dagger}(x=0,t)-D(x=0,t)c^{\dagger}],\nonumber\\
H_d&=&\ido (\omega-\omega_F) d^{\dagger}(\omega)d(\omega)
\eea
\section{Heisenberg equations of motion}
\subsection{Eliminating the continua}
For any operator $O$ that does not explicitly depend on time, we have the equation of motion
\be
\ddt O=i[H,O]
\ee
with $H$ the total Hamiltonian (\ref{H}).
We choose a time $t_0$ in the past at which we start the calculation (i.e., we solve the equations for $t>t_0$), and at which time the Heisenberg and Schr\"odinger picture operators are taken as equal. Those operators at that special time are our input operators and thus are also indicated by the subscript ``in.''

We first formally solve the equations for the continuum operators $b(\omega),g(\omega),$ and $d(\omega)$ and substitute those results into the equations of motion for arbitrary operators acting on the discrete quantum systems $a$, $F$ and/or $c$, thus eliminating the continua from the description.
For example, starting at the end, with modes $d(\omega)$ and the field operator $D(x)$, we obtain \cite{gardiner1985}
\be
D(x=0,t)=d_{{\rm in}}(t)+\sqrt{\Gamma}c(t)
\ee
with the ``free field'' given by
\be
d_{{\rm in}}(t)=\frac{1}{\sqrt{2\pi}}\ido
d_0(\omega)\exp(-i(\omega-\omega_F) (t-t_0)).
\ee
The operator $d_0(\omega):=d(\omega,t_0)$ is an initial value for $d(\omega,t)$ at time $t=t_0$.

For the field operator $G(x)$ we similarly obtain
\be
G(x=0,t)=g_{{\rm in}}(t)+\sqrt{\gamma_2}\ket{F_1}\bra{F_2}(t)
\ee
with the free field
\be
g_{{\rm in}}(t)=\frac{1}{\sqrt{2\pi}}\ido
g_0(\omega)\exp(-i\omega (t-t_0)).
\ee
For the field $B(x)$ which couples both to $a$ and to $F$ we find that at $x=0$ it contains two driving terms
\be
B(x=0,t)=b_{{\rm in}}(t)+
\sqrt{\kappa}a(t-\tau)+\sqrt{\gamma_1}\ket{F_0}\bra{F_1}(t)
\ee
with the free field
\be
b_{{\rm in}}(t)=\frac{1}{\sqrt{2\pi}}\ido
b_0(\omega)\exp(-i\omega (t-t_0)).
\ee
At the  location $x=-c\tau$ of the cavity we get just one driving term
\be
B(x=-c\tau,t)=b_{{\rm in}}(t)+
\sqrt{\kappa}a(t).
\ee
We can now write down the equations of motion for the operators corresponding to the three discrete quantum systems.
For example, for the cavity mode annihilation operator $a(t)$ we get 
\be\label{ddta}
\ddt a=-i\omega_a a-\sqrt{\kappa}[b_{{\rm in}}+
\frac{1}{2}\sqrt{\kappa}a].
\ee
(The ``extra'' factor of 1/2 on the r.h.s. comes from the use of $\int_{t_0}^t\!dt' \,\delta(t-t')a(t') =\frac12 a(t)$, where the delta function inside the integral comes from the approximation
$\int\!d\omega\, \exp(-i\omega(t-t'))=2\pi\delta(t-t')$ \cite{gardiner1985}. Instances of the same factor of 1/2 appear in several equations below.)\\
We can solve Eq.~(\ref{ddta}) to obtain
\bea
a(t)&=&a(t_0)\exp[(-i\omega_a-\kappa/2)(t-t_0)]
\nonumber\\
&&-\sqrt{\kappa}\int_{t_0}^t dt' \exp[(i\omega_a+\kappa/2)(t'-t)]b_{{\rm in}}(t').
\eea
For the evolution of $c$ we find
\be\label{ddtc}
\ddt c=-i\Delta c -\sqrt{\Gamma}(d_{{\rm in}}+\frac12\sqrt{\Gamma}c)-\tilde{F}.
\ee
The equation for system $F$ and hence for $\tilde{F}$ is more complicated and in general has to be solved numerically.
We can formally solve (\ref{ddtc})
 \bea\label{solvec}
c(t)&=&c(t_0)\exp[(-i\Delta-\Gamma/2)(t-t_0)]
\nonumber\\
&&-\int_{t_0}^t dt' \exp[(i\Delta+\Gamma/2)(t'-t)][\sqrt{\Gamma}d_{{\rm in}}(t')+\tilde{F}(t')].\nonumber\\
\eea
 and this is an explicit solution provided
 we can ignore the backaction of system $c$ on system $F$, i.e., when the operator $\tilde{F}(t)$ does not depend on ``downstream'' system $c$ operators, but only on ``upstream'' system $a$ operators. 
 \subsection{Steady-state solutions}
Our Hamiltonian is such that the $F$ system will reach a steady state,
The reason is that the force driving system $F$ is the single photon emitted by the cavity, and that photon will have disappeared after a few cavity life times $\kappa^{-1}$. The $F$ system then decays to either the $\ket{F_0}$ state or to the $\ket{F_2}$ state, and stays there.
The operator $\tilde{F}$ will eventually become constant, apart from fluctuating noise (Langevin) terms.
Eq. (\ref{solvec}) then shows that the operator $c$ will reach a steady state, too (all transient effects decay away at a rate $\Gamma$), up to noise terms.

We now focus on terms in $\tilde{F}$ proportional to $\proj{F_0}$ only. (These terms describe the response of molecule $F$ when it starts from state $\ket{F_0}$, where it is supposed to start. Other nonzero terms are discussed in Section \ref{num}.) 
We also assume $\delta=0$ for the moment (this is the optimum case, of course, for detecting the photon). 
Moreover, we set $F_0=F_1=0$ and
$F_2=:\mu>0$, so that the molecule triggers amplification only in the state $\ket{F_2}$. In that case
we simply have $\tilde{F}=\mu\proj{F_2}$, and we find its steady-state value to be
\bea\label{Fss}
\tilde{F}_{{\rm ss}}=\mu P_{\rm abs} a^{\dagger}a\otimes\proj{F_0},
\eea
where 
\bea
P_{\rm abs}=\frac{4\gamma_1\gamma_2}{(\gamma_1+\gamma_2)(\gamma_1+\gamma_2+\kappa)}
\eea
is the probability that the photon transfers the population from the initial state $\ket{F_0}$ to $\ket{F_2}$. This probability
 is maximized for $\gamma_1^2=\gamma_2^2+\kappa\gamma_2$, and
 for $\kappa\ll \gamma_{1,2}$ this maximum approaches unity arbitrarily closely. This result confirms the ``ideal detection'' result of Ref.~\cite{young2018}.
 
 We can in fact generalize this result to arbitrary detuning $\delta$. Apart from obtaining the answer by replacing $\kappa$ by $\kappa-2i\delta$ and taking the real part, 
 \be
 P_{\rm abs}=\Re\left[\frac{4\gamma_1\gamma_2}{(\gamma_1+\gamma_2)(\gamma_1+\gamma_2+\kappa-2i\delta)}\right]
 \ee
 we can write the result more insightfully in the form
 \be
  P_{\rm abs}=\int\!d\omega\,
  |\phi(\omega)|^2|T(\omega)|^2
 \ee
 where 
 \be\label{Tom}
 T(\omega)=\frac{\sqrt{\gamma_1\gamma_2}}{(\gamma_1+\gamma_2)/2-i(\omega-\omega_{10})}
 \ee
 is the transmission coefficient describing the transmission of a single excitation through the  $\Lambda$ system (which for a single excitation is equivalent to a Fabry-Perot filter cavity) \cite{propp2019,vanenk2017b} and where
 \be\label{phiom}
 \phi(\omega)=\frac{1}{\sqrt{2\pi}}\frac{\sqrt{\kappa}}{\kappa/2-i(\omega-\omega_a)}
 \ee
 is the (properly normalized) spectral shape of the photon produced by the cavity. This way of writing the probability can be generalized to other  systems than a three-level molecule by substituting other transmission functions $T(\omega)$ that describe the initial (absorption) stage of the photodetection process, as discussed in great detail in Ref.~\cite{propp2019}.
  
When the system reaches its steady state, the expression  for $c(t)$ becomes
\bea
c_{{\rm ss}}(t)=\tilde{d}(t)-\frac{\tilde{F}_{{\rm ss}}}{\Gamma/2+i\Delta},
\eea
where $\tilde{d}(t)$ is a single-mode ``noise'' annihilation operator given by (for large $t$, i.e., $t-t_0\gg 1/\Gamma$)
\bea\label{td}
\tilde{d}(t)=-\int_{t_0}^t dt' \exp[(i\Delta+\Gamma/2)(t'-t)]\sqrt{\Gamma}d_{{\rm in}}(t')
\eea
One can verify that 
\bea
[c_{{\rm ss}}(t),c^{\dagger}_{{\rm ss}}(t)]=1
\eea
thanks purely to the noise term.

What we observe in the end is a macroscopic amount of excitations in the continuum mode $d(\omega)$, or, equivalently, the field $D(x=0)$. The excitations are collected over some finite time interval of duration $T$ from $T_0$ to $T_0+T$ [much later than $t_0$] with some low efficiency $\eta$. We could assume that our signal is determined by
\be
S_D(T)=\eta\int_{T_0}^{T_0+T}\!dt\,\, \expect{D^{\dagger}(x=0,t)D(x=0,t)}=:\eta N_D(T),
\ee
which corresponds to collecting a fixed fraction $\eta$ of all excitations in the field $D$. We could also assume we collect data continuously as a function of $T$, by continuously monitoring the field $D$. Leaving out the noise terms we can write
\bea\label{NDT}
N_D(T)&=&\Gamma \int_{T_0}^{T_0+T}\!dt
\int_{t_0}^t\!d\tau \int_{t_0}^t\!d\tau'\,\,
\expect{\tilde{F}(\tau)\tilde{F}(\tau')}\times\nonumber\\
&&
\exp\left((-i\Delta+\Gamma/2)(\tau-t)+(i\Delta+\Gamma/2)(\tau'-t')\right).\nonumber\\
\eea
Here the expectation value $\expect{\tilde{F}(\tau)\tilde{F}(\tau')}$ must be calculated using the Quantum Regression Formula (or Theorem) \cite{carmbook}.

Alternatively we could assume  we collect a fraction of excitations in a particular single discrete time-integrated mode
\be
N'_D(T)=\eta \expect{d_T^{\dagger}d_T}
\ee
where, for example, when $\Delta=0$, we choose
\be\label{dTdef}
d_T=\frac{1}{\sqrt{T}}\int_{T_0}^{T_0+T}\!dt\,\,  D(x=0,t),
\ee
which similarly would contain $\expect{\tilde{F}(\tau)\tilde{F}(\tau')}$.
For either choice, the signal grows linearly with $T$ once $c$ reaches a steady state. We will focus on the  former choice in the numerical calculations, i.e., we assume that $N_D(T)$ contains our macroscopic signal.

Consider now the noise in our amplification process. We can write our discrete mode operator $d_T$ (in the steady state) as
\bea\label{dT}
d_{T,{\rm out}}=P_{\rm abs}\sqrt{G} (a^{\dagger}a)_{{\rm in}}\otimes (\proj{F_0})_{{\rm in}}+\tilde{e}_{{\rm in}},
\eea
where we explicitly added back in the subscripts ``out'' and ``in'' to indicate the operators on the right-hand side are all input operators and the left-hand side is an output operator. (Recall that we did leave out here other terms to be discussed in \ref{num}, given that the initial state of our molecule is $\ket{F_0}$.)
The gain factor $G$ here---which is the gain one gets if the molecule ends up in the desired state $ \ket{F_2}$---is linear in $T$
\bea\label{G}
G=\frac{4\mu^2}{\Gamma}T,
\eea
and $\tilde{e}_{{\rm in}}$ is a single-mode discrete annihilation noise operator fully determined by $d_{{\rm in}}$:
\bea
\tilde{e}_{{\rm in}}=\frac{1}{\sqrt{T}}\int_{T_0}^{T_0+T}\!dt\,\, 
(\sqrt{\Gamma}\tilde{d}(t)+d_{{\rm in}}(t)),
\eea
since the operator $\tilde{d}$ is determined by 
$d_{{\rm in}}$ according to Eq.~(\ref{td}).

Note that (i) the noise is not amplified, (ii) the first (gain) term is hermitian and, therefore, commutes with its hermitian conjugate, so that the presence of $\tilde{e}_{{\rm in}}$ is sufficient to preserve the commutator.
Of course, our operator $a^{\dagger}a$ is restricted here to the very narrow range of 0 or 1 excitations [so that we can use that $(a^{\dagger}a)^2=a^{\dagger}a$ when we calculate either $N_D(T)$ or $d^\dagger_T d_T$]. The main point is, Eq.~(\ref{dT}) is not of the Caves form for linear amplification, but is, rather, a nonlinear minimum-noise form that is akin to but different from the input-output relation found in Ref.~\cite{proppb}.
\subsection{Numerical integration}\label{num}
Without noise terms, we can find the gain term and the steady-state values numerically as well.
If the Schr\"odinger-picture evolution equation for the density operator
can be formally solved as $\rho(t)=\exp({\cal L}(t-t_0))\rho(t_0)$ with ${\cal L}$ the time-independent Liouvillian superoperator, then in the Heisenberg picture observable $O$ evolves as $O(t)=\exp({\cal L}^\dagger (t-t_0))O(t_0)$.

For the Heisenberg operator $\proj{F_2}(t)$ we plot all nonzero terms (there are five for $\delta=0$) as functions of time in Fig.~2.
Here are the  five types of terms with their interpretations (where we ignore operators acting on the Hilbert space for the ``downstream'' system $c$)
\begin{enumerate}
\item $K_1=\proj{1}\otimes \proj{F_0}$:
This term describes how an initial state with 1 cavity excitation and the molecule starting in $\ket{F_0}$
transfers the molecule to state $\ket{F_2}$ (blue curve).
\item 
$K_2=\proj{0} \otimes \proj{F_1}$: This term describes how the molecule reaches state $\ket{F_2}$ even without a photon present provided it starts in the upper state $\ket{F_1}$. It decays to $\ket{F_2}$ with probability 1/2, given that $\gamma_1=\gamma_2$ here (green curve).
\item 
$K_3=\proj{1} \otimes \proj{F_1}$: This term again corresponds to the molecule starting in the upper state $\ket{F_1}$, from which it decays to $\ket{F_2}$ with probability 1/2.  Initially it behaves like the previous case.
However, because of the presence of the photon, the molecule can also be transferred to the desired final state $\ket{F_2}$ by first decaying to $\ket{F_0}$ and then absorbing the photon (orange curve).
\item
$K_4=\openone \otimes \proj{F_2}$: This term describes the trivial case where the molecule starts in $\ket{F_2}$ and just stays there, independent of the presence or absence of a photon (dashed purple curve).
\item 
$K_5=a\otimes \ket{F_1}\bra{F_0}$+$a^\dagger \otimes \ket{F_0}\bra{F_1}$: This term describes the influence of coherence: if we start with a coherent superposition of no photon and 1 photon, {\em and} the molecule is in a superposition of $\ket{F_0}$ and $\ket{F_1}$, then the contributions from $\ket{0}\otimes\ket{F_1}$ and $\ket{1}\otimes\ket{F_0}$
to the
probabillity of ending up in $\ket{F_2}$ interfere destructively (dashed black/red curve).
Moreover, this term does contribute to the signal though the combination $K_5^\dagger K_5$, see main text.

(There is the similar sixth term, $K_6=ia\otimes \ket{F_1}\bra{F_0}$-i$a^\dagger \otimes \ket{F_0}\bra{F_1}$ which is nonzero only for nonzero detuning $\delta$.)
\end{enumerate}

\begin{figure}
\includegraphics[width=3in]{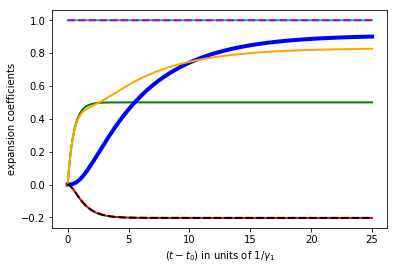}
\caption{The five types of nonzero terms in $\proj{F_2}(t)$ as functions of time in units of $\gamma_1^{-1}$,  where $\delta=0$, $\gamma_2=\gamma_1$ and $\kappa=\gamma_1/5$.
	The probability $P_{\rm abs}$ for the photon to trigger amplification is $P_{\rm abs}=10/11$ in this case. For details, see main text.
}
\end{figure}

\begin{figure}
\includegraphics[width=2.95in]{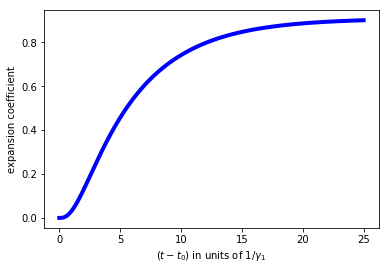}
\caption{The term in $c(t)$ (in units of $\mu/\Gamma$) proportional to
	$\proj{1}\otimes \proj{F_2} \otimes \proj{0}$ 
	which describes how the expectation value of the amplitude of our final quantum system $c$ grows with time (in units of $\gamma_1^{-1}$), if the cavity starts with 1 excitation, the molecule starts in $\ket{F_0}$ and the mode $c$ itself starts in the vacuum.
Parameter values are $\mu=\Gamma=\gamma_2=\gamma_1$, $\Delta=\delta=0$,  and $\kappa=\gamma_1/5$. The steady-state value of $c$ for these values is
$c_{\rm ss}=-20\mu/11\Gamma$.
}
\end{figure}

We also plot the amplitude of system $c$ and the total number of excitations in the field $D(x=0)$ in Figs.~3 and 4.  More precisely, we plot the time evolutions of the terms in $c(t)$ and $N_D(T)=D^\dagger(x=0) D(x=0)$ proportional to $\proj{1}\otimes\proj{F_0}\otimes\proj{0}$, which correspond to the system starting in the initial state with 1 excitation in the cavity, the molecule in state $\ket{F_0}$ and no excitations in mode $c$. We choose $T_0=t_0$ in the definition of $N_D(T)$. Note that it is not just the first term of the five terms we just discussed that contributes to $N_D(T)$: the fifth term contributes as well and so do the noise terms; together they ensure that $N_D(T)$ scales linearly with $P_{\rm abs}$ (as it should), rather than quadratically. (Numerically, we used the Quantum Regression Theorem to calculate $N_D(T)$.)
\begin{figure}
\includegraphics[width=2.35in]{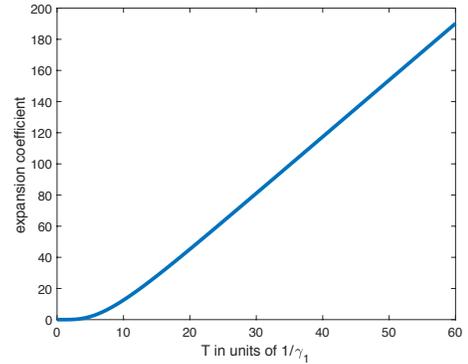}
\caption{The term in  $N_D(T)$, given by Eq.~(\ref{NDT}), proportional to $\proj{1}\otimes\proj{F_0}\otimes\proj{0}$ as a function of the integration time $T$ (in units of $\gamma_1^{-1})$. Parameter values are as in the previous Fig.
This represents the macroscopic signal produced by detection of  a single photon, i.e., the expectation value of the number of excitations in the field $D(x=0)$. This expectation value increases linearly with $T$ once the system has reached a steady state, after a few $\kappa^{-1}$, with a slope given by
$P_{\rm abs}4\mu^2 /\Gamma$.}
\end{figure}

The behavior of our signal plotted here and the efficiency $\eta$ with which that macroscopic signal is measured determine when the photodetector can be reset. Namely, as soon as a few excitations (in principle, even a single one, if we can ignore dark counts) have been detected (which requires the number of excitations to have been of order $1/\eta$),
a photon has been detected, and our detector may be reset in order to be able to detect a next photon.  Resetting involves emptying the system $c$ (which takes several $\Gamma^{-1}$) and resetting the molecule to the state $\ket{F_0}$ (which takes several $\gamma_1^{-1}$). 

If we include the Langevin terms we could solve the stochastic differential equations by standard methods (using Ito calculus, for example \cite{gardiner1985}). Much more simply, we could determine what the generic form of the noise terms must be, by making use of the fact that commutators for our output operators like $d_T$ should be preserved. For example, if the steady-state value of $d_T$ is $d_{T,{\rm ss}}$ (which is expressed in terms of the input operators of our discrete systems) and the noise term in $d_T$ is $e_{{\rm in}}$ (which is expressed in terms of continuum input operators and which, therefore, commutes with $d_{T,{\rm ss}}$ and $d^\dagger_{T,{\rm ss}}$), then we  must have
\be
[e_{{\rm in}},e^\dagger_{{\rm in}}]=\openone-[d_{T,{\rm ss}},d^\dagger_{T,{\rm ss}}],
\ee
which limits the operator form for $e_{{\rm in}}$ severely.
\section{Conclusions and outlook}\label{discuss}
We used a fairly simple Hamiltonian model to describe quantum mechanically the photodetection process from beginning (generation and absorption of a single photon) to end (amplification). We solved the equations in the Heisenberg picture, because the end result compactly describes the essence of the whole process including the noise therein (see Eq.~(\ref{dT}), where we  consider the optimum case $P_{\rm abs}=1$):
\be
d_{{\rm out}}=\sqrt{G}(a^\dagger a)_{\rm in}\otimes \proj{F_0}
+e_{\rm in}.
\ee
Here $d_{{\rm out}}$ is the annihilation mode operator for a time-integrated mode that contains the macroscopic output signal (a large number of excitations) that we ultimately observe classically. We have to place the Heisenberg cut somewhere, and we place it as far along the whole photodetection process as we can, {\em after amplification}. Unlike for linear amplification \cite{caves1982} where the gain term would be $\sqrt{G}a_{\rm in}$, here the gain term indicates the amplification process is nonlinear \cite{proppb}.
The noise term (needed to preserve the commutator $[d,d^\dagger]=\openone$) is just an input mode  operator, rather than $\sqrt{G-1}e^\dagger$, which is the noise term accompanying phase-insensitive linear amplification. That is, the noise in our case is not amplified, and a vacuum input gives zero noise, unlike for linear amplification.
The projector $\proj{F_0}$ projects onto the initial state of a molecule that triggers the amplification process once it has changed its configuration by absorbing the photon. Confirming results of Refs.~\cite{young2018,young2018b} the probability of absorption (and detection) can indeed be nearly 100\%.

Several aspects of our minimal prototype Hamiltonian can be generalized and/or extended:

(i) In order to detect more than a single excitation we could include more levels in the $F$ molecule (and more excitations either in the same cavity or in multiple cavities, all coupled to the same system $F$).
For example, to be able to detect a second photon we could introduce two more $F$ levels, such that a transition from $\ket{F_2}$ to another upper level could occur (triggered by the second photon), which then could decay to a level $\ket{F_4}$ where the value of the parameter $F_4$ would be substantially larger than $F_2=\mu$.  This then would allow us to distinguish the signal produced by the second photon from that produced (triggered) by just a single photon. 

(ii) In \cite{yang2019} a quantum phase transition was proposed and analyzed as a means of amplifying a weak signal (such as a single photon).
Here we could use a {\em dissipative} phase transition \cite{dimer2007,kessler2012,carmichael2015,fink2017} to achieve minimum-noise amplification. The Hamiltonian $H_c(F)$ we used is for either a driven atom or a driven cavity. A dissipative phase transition arises even for the simple system of an atom inside (and coupled to) a cavity, with either the atom or the cavity driven. The presence of a phase transition may make the amplification process more robust against deviations from the ideal Hamiltonian.

(iii) The single-photon wavepacket to be detected is fixed here by the resonance frequency and the decay rate of the cavity that generates the photon. We could make these two parameters arbitrary functions of time so that an arbitrary single-photon wavepacket can be created \cite{gheri1998,molmer2019,molmer2020}. That should allow us to formulate the POVM (projecting onto a specific temporal state of the photon) that describes our detector, as in  Ref.~\cite{propp2020}.

(iv) We assumed a {\em bosonic} mode to contain the amplified signal.
Alternatively we may use many spin-1/2 particles, as in the model discussed in Refs.~\cite{yang2019,yang2019b}. This extension would increase the scope of our description to include fermionic systems.

In conclusion, the main point here was to present a class of Hamiltonians
that describe the  photodetection process fully quantum-mechanically from beginning to end, including nonlinear, minimum-noise amplification \cite{proppb} and near-perfect photoabsorption 
\cite{young2018}.

\section*{Acknowledgments}
We thank Tzula Propp for their useful comments and discussions.

This work is supported by funding from
DARPA under
Contract No. W911NF-17-1-0267.

\bibliography{Heisenberg}
\end{document}